\documentclass[11pt]{iopart}
 \usepackage[active]{srcltx}
 \usepackage{amsthm,graphicx,amsfonts,amssymb,latexsym,cancel,color,amsbsy,ytableau,tikz,setstack}
  \eqnobysec

  \theoremstyle{plain}
 \newtheorem {hypo}{\bf\hspace{-\parindent}Hypothesis}

 \newtheorem {prop}[hypo]{Proposition}
 
 \newtheorem {conj}[hypo]{Conjecture}
 
 \theoremstyle{definition}

 \theoremstyle{remark}

 \newcommand\Zb{\mathbb{Z}}
 \newcommand\Rb{\mathbb{R}}
 \newcommand\Cb{\mathbb{C}}
 \newcommand\Pb{\mathbb{P}}

 \newcommand\ben{\begin{equation*}}
 \newcommand\ebn{\end{equation*}}
 \newcommand\be{\begin{equation}}
 \newcommand\eb{\end{equation}}

 \newcommand{\pf}{\begin{bpf}}

 \newcommand{\pfms}{\begin{bpfms}}
 \newcommand{\epf}{\end{bpf}\hfill$\square$\vspace{0.1cm}}
 \newcommand{\epfms}{\end{bpfms}\hfill$\square$\\ }

\begin{document}
 \title[]{Connection problem for the sine-Gordon/Painlev\'e III tau function and irregular conformal blocks}
 \author{A. Its$^1$, O. Lisovyy$^2$, Yu. Tykhyy$^3$}
 
 \address{$^1$ Department of Mathematical Sciences,
 Indiana University-Purdue University,
 402~N.~Blackford St.,
 Indianapolis, IN 46202-3267,
 USA}
 \address{$^2$ Laboratoire de Math\'ematiques et Physique Th\'eorique CNRS/UMR 7350,
  Universit\'e de Tours,
  37200 Tours, France}
 \address{$^3$ Bogolyubov Institute for Theoretical Physics,
 03680, Kyiv, Ukraine}
  \eads{\mailto{itsa@math.iupui.edu},  \mailto{lisovyi@lmpt.univ-tours.fr}, \mailto{tykhyy@bitp.kiev.ua}}

 \begin{abstract}
 The short-distance expansion of the tau function of the radial sine-Gor\-don/Painlev\'e III equation is given by  a convergent series which involves irregular $c=1$ conformal blocks and possesses certain periodicity properties with respect to monodromy data. The long-distance irregular expansion  exhibits a similar pe\-ri\-o\-dicity with respect to a different pair of coordinates on the monodromy manifold. This observation is used to conjecture an exact expression for the connection constant  providing relative normalization of the two series. Up to an elementary prefactor, it is given by the gene\-rating function of the canonical transformation between the two sets of coordinates.
 \end{abstract}
 
 \section{Introduction}
 Since the late seventies, there
 has been an explosive growth in the application of Painlev\'e equations to an
 extraordinary variety of problems, both mathematical and physical. 
 Of particular importance is the role that the Painlev\'e functions play in problems related to random matrices and
 random processes, orthogonal polynomials,  string theory, and in exactly solvable statistical mechanics and
 quantum field models. 
 
 Most recently, Painlev\'e transcendents  have been understood to play a role  in the two-dimensional
 conformal field theory \cite{GIL,ILT,Lit,T11}. 
 This connection produces, in parti\-cular,  a new type of   representation of
 the Painlev\'e functions in the form of explicit  series which  have a meaning of the Fourier transforms of the $c=1$ Liouville conformal blocks and their irregular counterparts \cite{GIL,GIL2}.  It also  proved to be 
 very useful in tackling one of the most difficult problems of the analytic theory of Painlev\'e functions, the problem of evaluation  of the 
 constant factors in the asymptotics of  the Painlev\'e tau functions  \cite{ILTy}.
 
 Usually, it is not the Painlev\'e  functions {\it per se} but the related {\it tau functions} 
 that are objects which actually appear in applications, notably in the description of the correlation functions of integrable statistical
 mechanics and quantum field models. The main analytic issue in these applications is the large time and
 distance  asymptotics with a particular focus on the relevant connection formulae between different 
 critical expansions and the evaluation of the above mentioned constant factors. The latter, very often, contain
 the most important information of the models in question. 
 Starting from the seminal works of Onsager and Kaufman on the Ising model whose 
 mathematical needs led to the birth of the Strong Szeg\"{o} Theorem
 in the theory of Toeplitz matrices (see e.g. \cite{DIKb} for more on the history of the matter), the evaluation of
 the constant terms in the asymptotics of different correlation and distribution functions of the random matrix
 theory and of the theory of solvable statistical mechanics models has always  been
 a great  challenge in the field.   
 In addition to the  Strong Szeg\"{o} Theorem we mention
 the work of Tracy \cite{T}, where the ``constant problem'' related to
 the Ising model was solved. This was the first rigorous solution of a ``constant problem'' for Painlev\'e equations (a special Painlev\'e~III transcendent).  Further developments \cite{K}, \cite{E}, \cite{DIKZ} were devoted to the rigorous derivation of Dyson's constant \cite{dyson} in the asymptotics of the sine-kernel determinant describing the gap probability in large random matrices. This determinant represents 
 the tau function of a particular solution of the fifth Painlev\'e equation. Other ``constant'' problems were also considered  and solved in the works \cite{BT}, \cite{BB}, \cite{DIKa}, \cite{Lis} and \cite{BBDiF}.

 A natural framework for the global asymptotic analysis of the solutions of Painlev\'e equations is
 provided by the Isomonodromy-Riemann-Hilbert method, see monograph \cite{FIKN}
 for more detail and for the history of the subject. However, the Riemann-Hilbert technique
 addresses directly the Painlev\'e functions and not the associated tau functions. The latter  are
 logarithmic antiderivatives of certain rational functions of Painlev\'e transcendents --- 
 the Hamiltonians of Painlev\'e equations. Hence the problem: one should be able to
 evaluate  integrals of  certain combinations of Painlev\'e transcendents and their
 derivatives. So far, this problem has been successfully  handled only for very special 
 solutions of the Painlev\'e equations. The tau functions evaluated on these solutions
 admit additional representations in terms of certain Fredholm or Toeplitz determinants.
 It is this extra property that, in spite of the  difference in the approaches, is a principal reason of the 
 success in the solution of the ``constant problem'' in all of the works mentioned in the previous paragraph.

 The situation with the ``constant problem'' has changed very recently. Conformal block representations of the generic Painlev\'e VI tau function  \cite{GIL} imply that the con\-nec\-tion constants satisfy certain functional relations. They have been solved in \cite{ILTy} using the hamiltonian interpretation of the monodromy parametrization of  Painlev\'e VI trans\-cen\-dents
  provided via the Isomonodromy-Riemann-Hilbert framework. Indeed, as every other 
 Painlev\'e equation, the sixth Painlev\'e equation  is a classical hamiltonian system,
 and the monodromy data form a canonically conjugated pair of  coordinates on the space of its solutions. The parametrization of the asymptotics of the tau function  in terms of monodromy 
 data implies in turn that the correspondence  between the asymptotic parameters at the different critical
 points can be interpreted as a canonical map between different sets of canonical coordinates. The
 asymptotic constants in question can be thought of as the {\it generating functions} of these maps. 
 This observation has allowed to formulate detailed conjectures about the dependence of the Painlev\'e VI asymptotic  constants on the monodromy data \cite{ILTy}, i.e., practically, to evaluate
 them in explicit form.
 
 In the present paper we extend the method of \cite{ILTy} to a special case of the third Painlev\'e 
 equation, denoted as Painlev\'e III$_3$ or P$_{\rm III}\left(D_8\right)$ \cite{ohyama}, 
 \begin{equation}\label{rSG0}
   u_{rr}+\frac{u_r}{r}+\sin u=0,
 \end{equation}
 which appears in physical applications as a reduction of the sine-Gordon equation.  
 
 Following the definitions of \cite{JMU}, the Painlev\'e III$_3$ tau function   is given by the formula
  \begin{equation}\label{taudef0}
  \frac{d}{dr}\ln \tau\left(2^{-12}r^4\right)
  =\frac{r}{4}\left[\left(\frac{iu_r}{2}+\frac1r\right)^2+\frac{\cos u}{2}\right].
  \end{equation}
  
  We are concerned with two representations of the tau function. The first one is its short-distance ($r\rightarrow0$) expansion. This expansion has been known already \cite{GIL2,ILST} and is given by the convergent series
  \be\label{tauexp00}
   \tau(t)=\sum_{n\in\Zb}e^{4\pi i n \eta}\mathcal{F}\left(\sigma+n,t\right),
  \eb
  where the function $\mathcal{F}\left(\sigma,t\right)$ is the irregular $c=1$ Virasoro conformal block 
  normalized as indicated in the equations (\ref{irrcb1})--(\ref{irrcb2}) below.
  The parameters $\sigma$ and $\eta$ are related to the mo\-no\-dromy of the associated linear system, see Section~\ref{sec_lsys}
  for their exact definition.
  The AGT correspondence identifies  $\mathcal{F}\left(\sigma,t\right)$ with an instanton partition 
  function given by a sum over pairs of Young diagrams \cite{Nekrasov,NO}, see formulae (\ref{irrcb3})--(\ref{irrcb4}). One of our results (Pro\-po\-sition~\ref{prop1}) is the proof of convergence of this sum and the series (\ref{tauexp00}).
 
 The second representation corresponds to the long-distance ($r\rightarrow\infty$) expansion. We suggest that the {\it same} function $\tau(t)$ can be repre\-sented as
 \begin{equation} \label{tauexpi10}
  \tau\left(2^{-12}r^4\right)=\chi\left(\sigma,\nu\right)\sum_{n\in\mathbb{Z}}e^{4\pi i n \rho}
  \mathcal{G}\left(\nu+in,r\right).
 \end{equation}
 The structure and normalization
 of the function  $\mathcal{G}\left(\nu,r\right)$ is described in the equations (\ref{tauexpi2})--(\ref{tauexpi3}) of the main text.  The parameters $\nu$ and $\rho$ 
 are explicit elementary functions of $\sigma$ and $\eta$, see (\ref{ldmndr1}) and (\ref{rhonot}). The series (\ref{tauexpi10}) is our {\it first conjecture} regarding the structure of 
 the long-distance asymptotics of the Painlev\'e III$_3$ tau function. It is based
 on the analysis of several first terms in the long-distance asymptotics of the
 function $u(r)$ which are available via the Riemann-Hilbert method.  The irregular expansion (\ref{tauexpi10}) 
  suggests the existence of novel distinguished bases in spaces of irregular conformal blocks \cite{GT}. An important open problem would be to understand the representation-theoretic origin of such bases.
 
 The quantity $\chi\left(\sigma,\nu\right)$, independently of the conjecture (\ref{tauexpi10}), has a well-defined meaning of the constant prefactor in the long-distance asymptotics of the tau function.
 Our {\it second conjecture}
 concerns this  prefactor. The periodicity of asymp\-totic expansions around both singularities  enables one to develop a hamiltonian-based scheme similar to  Painlev\'e~VI arguments outlined above.
 This scheme produces an explicit conjectural expression for the constant $\chi\left(\sigma,\nu\right)$ (equation (\ref{conn}) in the end of the paper), which is further confirmed by numerics. In  our future work we hope to be able to prove this conjecture
 using the theory of non-isomonodromy deformations of the
 tau functions of integrable models, recently developed
 in \cite{Bert}.

 We want to conclude this introduction by another remark on the structure of the expansions (\ref{tauexp00}) and (\ref{tauexpi10}). 
 The periodicity with respect to monodromy data is completely lost in the  leading terms of the asymptotics of the function 
 $u(r )$ at $r = 0$ and at $r =\infty$. The series (\ref{tauexp00}) and (\ref{tauexpi10}) tell us that  full asymptotic expressions can 
 be obtained by summing up all the integer shifts of the  leading terms. More precisely, in order for this procedure to work, one has
 to switch from the Painlev\'e function $u(r )$ to the tau function $\tau(t)$ and replace the 
 leading term of the asymptotics by the appropriate conformal block. It is interesting to note that very similar
 effect of  ``loosing''  the periodicity with respect to the relevant parameter and its ``recovering''  
 by summing up over all the integer shifts  had already emerged in the 1991's Basor-Tracy 
 conjecture \cite{BT1}  concerning the general Fisher-Hartwig asymptotics in the theory of Toeplitz determinants. 
 For more details, we refer the reader to \cite{DIK}  where the
 Basor-Tracy conjecture  was proven and \cite{Mir} where it was  used in a concrete application.

 \vspace{0.1cm}
 \noindent
 {\small {\bf Acknowledgements}. We are grateful to N. Iorgov, V. Novokshenov and J. Teschner for helpful discussions. The present work  was supported by the  NSF Grant DMS-1001777, Ukrainian SFFR project F54.1/019 and the IRSES project ``Random and integrable models in mathematical physics''}.

 \section{Background}\label{sec_lsys}
 \subsection{Auxiliary linear problem}
 Let us briefly outline the relation of the radial sine-Gordon equation to
 the theory of monodromy preserving deformations of linear ODEs with rational coefficients. The reader is referred
 to \cite{FIKN,Niles} for more details.

 Consider the following linear system:
 \begin{eqnarray}\label{lsys1}
 \frac{\partial \Phi}{\partial z}=A\left(z\right)\Phi,\\
 A\left(z\right)=-\frac{ir^2\sigma_3}{16}-\frac{iv\sigma_1}{4z}+\frac{ie^{-\frac{iu\sigma_1}{2}}\sigma_3
 e^{\frac{iu\sigma_1}{2}}}{z^2}.
 \end{eqnarray}
 Here $\Phi(z)$ is a multivalued $2\times 2$ matrix function on $\Pb^1\backslash\{0,\infty\}$, the scalars $r,u,v$
 are independent of $z$, and
 $\sigma_{1,2,3}$ denote the Pauli matrices,
 \ben
 \sigma_1=\left(\begin{array}{cc} 0 & 1 \\ 1 & 0 \end{array}\right),\qquad
 \sigma_2=\left(\begin{array}{cc} 0 & -i \\ i & 0 \end{array}\right),\qquad
 \sigma_3=\left(\begin{array}{cc} 1 & 0 \\ 0 & -1 \end{array}\right),
 \ebn
 The system (\ref{lsys1}) has two irregular singular points $z=0,\infty$ of Poincar\'e rank $1$. The formal fundamental solutions
 around these points have the form
 \begin{eqnarray}
 \label{as0}
 \Phi^{(0)}(z)&=e^{-\frac{iu\sigma_1}{2}}
 \left[1+\sum_{k=1}^{\infty}\phi_{k}^{(0)}z^{k}\right]e^{-\frac{i\sigma_3}{z}},\\
 \label{asi}
 \Phi^{(\infty)}(z)&=\left[1+\sum_{k=1}^{\infty}\phi_{k}^{(\infty)}z^{-k}\right]e^{-\frac{ir^2\sigma_3 z}{16}}.
 \end{eqnarray}
 The asymptotics (\ref{as0})--(\ref{asi}) uniquely determines the canonical solutions $\Phi^{(0,\infty)}_{1,2,3}(z)$ in the Stokes sectors $\mathcal{S}^{(0,\infty)}_{1,2,3}$ defined by
 \begin{eqnarray*}
 \mathcal{S}^{(0)}_{k}=&\left\{z: (k-2)\pi<\mathrm{arg}\,z< k\pi,\; |z|< R\right\},\\
 \mathcal{S}^{(\infty)}_{k}=&\left\{z: \left(k-\frac32\right)\pi-2\epsilon<\mathrm{arg}\,z< 
 \left(k-\frac12\right)\pi+2\epsilon,\; |z|> R\right\},
 \end{eqnarray*}
 with $k=1,2,3$ and small finite $\epsilon>0$.

 \subsection{Monodromy data}
 The set of monodromy data consists of four Stokes matrices
 \ben
 S_{k\rightarrow k+1}^{(p)}={\Phi_{k}^{(p)}}^{-1}\left(z\right)\Phi_{k+1}^{(p)}\left(z\right),\qquad p=0,\infty,\quad k=1,2,
 \ebn
 and one connection matrix
 \ben
 C={\Phi_{1}^{(0)}}^{-1}\left(z\right)\Phi_{1}^{(\infty)}\left(z\right).
 \ebn
 The familiar triangular structure of the Stokes matrices and the symmetry
 $\sigma_1 A\left(z\right)\sigma_1=-A\left(-z\right)$ of the  linear system (\ref{lsys1}) imply
 that
 \begin{eqnarray}\label{stokes1}
 S_{1\rightarrow 2}^{(0)}=\sigma_1 S_{2\rightarrow 3}^{(0)}\sigma_1=
 \left(\begin{array}{cc} 1 & a \\ 0 & 1\end{array}\right),\\
 \label{stokes1v2}
 S_{1\rightarrow 2}^{(\infty)}=\sigma_1 S_{2\rightarrow 3}^{(\infty)}\sigma_1=
 \left(\begin{array}{cc} 1 & 0 \\ b & 1\end{array}\right).
 \end{eqnarray}
 The same symmetry constraints the form of the connection matrix. Indeed, we have the following relations
 \be\label{auxrels1}
 \sigma_1 C\sigma_1=\left(S_{1\rightarrow 2}^{(0)}\right)^{-1} C \,S_{1\rightarrow 2}^{(\infty)}=
 S_{2\rightarrow 3}^{(0)}\, C \left(S_{2\rightarrow 3}^{(\infty)}\right)^{-1}.
 \eb
 Also, since $\mathrm{det}\,\Phi(z)$ is independent of $z$,
 the normalization of the formal solutions (\ref{as0})--(\ref{asi}) implies that
  $\mathrm{det}\,C=1$. Now it follows from (\ref{auxrels1})  that
  in the generic case the connection matrix can be parameterized as
  \be\label{connm}
  C=\frac{1}{\sin 2\pi\sigma}\left(\begin{array}{cc}
  \sin2\pi\eta & -i\sin2\pi(\eta+\sigma) \\ i\sin2\pi(\eta-\sigma) & \sin2\pi\eta \end{array}\right),
  \eb
  and, moreover, the Stokes factors $a,b$ in (\ref{stokes1})--(\ref{stokes1v2}) are given by
  \be\label{stokes2}
  a=b=-2i\cos2\pi\sigma.
  \eb
  Thus the pair $\left(\sigma,\eta\right)\in\Cb^2$ determines the monodromy of the linear system (\ref{lsys1}).
  Since the connection matrix and Stokes factors remain invariant under the transformation
  $(\sigma,\eta)\rightarrow(-\sigma,-\eta)$,
  it can be assumed without loss of generality that $0\leq\Re\sigma\leq \frac12$ and $ -\frac12<\Re \eta\leq\frac12$.

  \subsection{Monodromy preserving deformation}
  The above construction defines the monodromy map
  $\mathfrak{m} $
  from the parameter set $\mathcal{P}$ of the linear system (\ref{lsys1}) to the moduli space $\mathcal{M}$ of monodromy data.
  The triple $\left(r,u,v\right)$ and the pair $\left(\sigma,\eta\right)$ can be seen as local coordinates on these
  two spaces.

  Suppose that $(r,u,v)$ vary in such a way that the monodromy remains constant. It is convenient
  to consider $r$ as a parameter and $u$, $v$ as smooth functions of $r$. The isomonodromy condition
  implies that $\displaystyle \frac{\partial\Phi}{\partial r}\,\Phi^{-1}$ is a meromorphic function on $\Pb^1$. In fact, from
  (\ref{as0})--(\ref{asi}) one obtains
  \begin{eqnarray}\label{lsys2}
  \frac{\partial \Phi}{\partial r}=B\left(z\right)\Phi,\\
 B\left(z\right)=-\frac{ir\sigma_3}{8}\,z-\frac{iu_r\sigma_1}{2}.
  \end{eqnarray}
  
  Matrix equations (\ref{lsys1}) and (\ref{lsys2}) provide a Lax pair for the radial sine-Gordon equation. Indeed,
  the zero-curvature condition $\left[\partial_z-A,\partial_r-B\right]=0$ is equivalent to
  two ODEs:
  \begin{eqnarray}
  \label{rSGaux}
   v=ru_r,\\
  \label{rSG}
  u_{rr}+\frac{u_r}{r}+\sin u=0.
  \end{eqnarray}
  These equations  can be rewritten as a non-autonomous Ha\-mil\-tonian system
  \begin{equation*}
  u_r=\frac{\partial\mathcal{H}}{\partial v}, \qquad 
  v_r=-\frac{\partial\mathcal{H}}{\partial u},
  \end{equation*}
  with the Hamiltonian given by
  \be
  \label{hamSG}
  \mathcal{H}=\frac{v^2}{2 r}-r\cos u.
  \eb
  
  Note that the monodromy parameters $(\sigma,\eta)$ can be interpreted as two integrals of motion for
  the sine-Gordon equation (\ref{rSG}). The formulae for the inverse monodromy map $\mathfrak{m}^{-1}:\left(\sigma,\eta\right)\mapsto
  u(r)$ in terms of explicitly defined series will be provided
  in Section~\ref{sec_cr}.

  \subsection{Tau function and relation to Painlev\'e III}
  The tau function of the radial sine-Gordon equation (\ref{rSG}) will be defined by
  \begin{eqnarray}\label{taudef}
  \frac{d}{dr}\ln \tau\left(2^{-12}r^4\right)&=\frac{r}{4}\left[\left(\frac{iu_r}{2}+\frac1r\right)^2+\frac{\cos u}{2}\right]=\\
  \nonumber
  &= -\frac{\mathcal{H}}{8}+\frac14\frac{d}{dr}\ln r e^{iu}.
  \end{eqnarray}
  Its logarithmic derivative $\displaystyle \zeta(t)=t\frac{d}{dt}\ln\tau(t)$ satisfies the $\sigma$-form of Painlev\'e~III$_3$:
  \be\label{spiii3}
  \left(t\zeta''\right)^2=4\left(\zeta'\right)^2\left(\zeta-t\zeta'\right)-4\zeta'.
  \eb
  Also, setting $\displaystyle s=2^{-6} r^2$, $q(s)=-e^{iu(r)}$, we get Painlev\'e~III$_3$ in the standard form,
  \be
  \label{piii3}
  q''=\frac{\left(q'\right)^2}{q}-\frac{q'}{s}+\frac{8\left(q^2-1\right)}{s}.
  \eb
  Converse relations between $u$, $\tau$ and $\zeta$ are given by
  \be\label{convrels}
  e^{-iu(r)}=4r^{-1}\frac{d}{dr} r\frac{d}{dr}\ln \tau\left(2^{-12}r^4\right)=2^{-6}r^2\,\zeta'\left(2^{-12}r^4\right).
  \eb
  It is important to emphasize that (\ref{taudef}) determines the tau function up to a constant factor. In applications,
  however, there is often a distinguished normalization coming from the physical context which fixes this ambiguity.

  \section{Critical expansions}\label{sec_cr}
  \subsection{Short-distance expansion}
  Let us assume that $-\frac{\pi}{4}+\epsilon <\arg r<\frac{\pi}{4}-\epsilon$
  and consider generic situation where the Stokes parameter $\sigma$ belongs to the strip
  $0<\Re\sigma <\frac12$.  The asymptotic behavior of $u(r)$ as
  $r\rightarrow 0$ is then given by \cite{Ji,Nov,IN}
  \be\label{shas}
  e^{iu(r)}=-e^{4\pi i \eta}\frac{\Gamma^2\left(1-2\sigma\right)}{\Gamma^2\left(2\sigma\right)}\left(\frac{r}{8}\right)^{8\sigma-2}\bigl[1+o(1)\bigr],
  \eb
  see also \cite{Niles} and \cite[Chapter 14]{FIKN}.
  Subleading corrections may be calculated using the following ansatz:
  \be\label{shas2}
  u(r)=\alpha\ln r+\beta+\sum_{k=1}^{\infty}\sum_{l<k}\sum_{\epsilon=\pm}
  a_{kl}^{\epsilon}r^{2k+i\epsilon l \alpha}.
  \eb
  Here the first coefficients $\alpha,\beta$ are read off the asymptotics (\ref{shas}), whereas  $a^{\pm}_{kl}$ can be determined recursively from the sine-Gordon equation.

  It was discovered very recently \cite{GIL2,ILST} that the whole short-distance expansion can be written down \textit{explicitly}
  if one uses the tau function instead of $u(r)$. The result is given by a Fourier transform
  \be\label{tauexp0}
  \tau(t)=\sum_{n\in\Zb}e^{4\pi i n \eta}\mathcal{F}\left(\sigma+n,t\right),
  \eb
  of the irregular $c=1$ Virasoro conformal block  $\mathcal{F}\left(\sigma,t\right)$ normalized as
  \begin{eqnarray}
  \label{irrcb1}
  \mathcal{F}\left(\sigma,t\right)=\frac{t^{\sigma^2}\mathcal{B}\left(\sigma,t\right)}{
  G\left(1+2\sigma\right)G\left(1-2\sigma\right)},\\
  \label{irrcb2}
  \fl\qquad \mathcal{B}\left(\sigma,t\right)=
  1+\frac{t}{2\sigma^2}+\frac{\left(8\sigma^2+1\right)t^2}{4\sigma^2\left(4\sigma^2-1\right)^2}+
 \frac{\left(8\sigma^4-5\sigma^2+3\right)t^3}{24\sigma^2\left(\sigma^2-1\right)^2\left(4\sigma^2-1\right)^2}+\ldots
  \end{eqnarray}
  Here $G(z)$ denotes the Barnes $G$-function, which satisfies the functional equation $G(z+1)=\Gamma(z)G(z)$. The coefficients of the series (\ref{irrcb2}) are rational functions of $\sigma^2$; their
  poles $\sigma\in\mathbb{Z}/2$ correspond to zeros of $c=1$ Kac determinant.

  Irregular conformal blocks involve coherent states (Whittaker vectors)
  on which the annihilation part of the Virasoro algebra acts diagonally \cite{Gaiotto1,GT}. We are dealing here with
  one of the simplest cases where conformal block coincides with the norm of such state,
  $\mathcal{F}\left(\sigma,t\right)={}_W\langle \sigma|\sigma\rangle_W$, satisfying
  \begin{eqnarray*}
  L_0|\sigma\rangle_W=2t\,\partial_t |\sigma\rangle_W,\\ L_1 |\sigma\rangle_W=\sqrt{t}\,|\sigma\rangle_W,
  \\ L_{n\geq 2}|\sigma\rangle_W=0.
  \end{eqnarray*}
  Painlev\'e III$_3$ independent variable is thus related to the only nontrivial eigenvalue.

  AGT duality relates conformal block $\mathcal{F}\left(\sigma,t\right)$ to the partition function
  of the  $\mathcal{N}=2$ supersymmetric pure $SU(2)$ gauge theory.
  This allows to write the series in (\ref{irrcb2}) as a Nekrasov instanton
  sum \cite{Nekrasov,NO} over pairs of Young diagrams:
  \begin{eqnarray}
  \label{irrcb3}\mathcal{B}\left(\sigma,t\right)=
  \sum_{\lambda,\mu\in\mathbb{Yb}}
  \left(\frac{\dim\lambda\dim\mu}{|\lambda|!\,|\mu|!} \right)^2 \frac{t^{|\lambda|+|\mu|}}{\left[b_{\lambda,\mu}\left(\sigma\right)\right]^2},\\
  \label{irrcb4}
  \fl\qquad
  b_{\lambda,\mu}\left(\sigma\right)=\prod_{(k,l)\in\lambda} \left(\lambda'_l-k+\mu_k-l+1+2\sigma\right)
  \prod_{(k,l)\in\mu} \left(\mu'_l-k+\lambda_k-l+1-2\sigma\right).
  \end{eqnarray}
  Here the diagrams $\lambda,\mu$ are identified with partitions by $\lambda=\{\lambda_1\geq \lambda_2\geq\ldots\geq \lambda_{\ell(\lambda)}>0\}$, the size of $\lambda\in\mathbb{Y}$ is denoted by $|\lambda|=\sum_{k=1}^{\ell(\lambda)} \lambda_k$, and  $\lambda'$ corresponds to the transposed diagram. $\dim\lambda$ denotes the dimension of the irreducible
  representation of the symmetric group $S_{|\lambda|}$ associated to $\lambda$. It can be calculated
  using the hook-length formula
  \ben
  \frac{\dim\lambda}{|\lambda|!}  =\frac{1}{\sqrt{b_{\lambda,\lambda}\left(0\right)}}.
  \ebn

  The formulae (\ref{tauexp0})--(\ref{irrcb4}) provide a series solution to the inverse monodromy problem for the radial sine-Gordon/Painlev\'e
  III$_3$ equation. They were obtained in \cite{GIL2,ILST} as a limiting case of a similar statement for Painlev\'e VI equation
  \cite{GIL}. Its CFT derivation and further generalizations were suggested in \cite{ILT}. The approach of \cite{ILT} is based on the braiding/fusion transformations of the Virasoro conformal blocks with degenerate fields. An alternative representation-theoretic proof of
  (\ref{tauexp0}) was recently
  found by M. Bershtein and A.~Shchechkin \cite{BSh}.

  Conformal blocks are usually considered as formal multivariate power series. However, in the case we are interested in their
  meaning can be made more precise.

  \begin{prop}\label{prop1} Let $2\sigma\notin\Zb$. Then:
  \begin{enumerate}
  \item conformal block series
  (\ref{irrcb3})--(\ref{irrcb4}) converges uniformly and absolutely on every bounded subset of  $\Cb$,
  \item tau function series (\ref{tauexp0}) converges uniformly and absolutely on every bounded subset of the universal cover of $\Cb\backslash\{0\}$.
  \end{enumerate}
  \end{prop}
  \pf Introduce the notation $L=\min_{n\in\Zb}\left|2\sigma-n\right|^2>0$. Then already  the roughest estimate
  $\left|b_{\lambda,\mu}(\sigma)\right|^2\geq L^{|\lambda|+|\mu|}$ implies that
  \ben
  \left|\mathcal{B}\left(\sigma,t\right)\right|\leq \left[\sum_{\lambda\in\mathbb{Y}}\left(\frac{\dim \lambda}{|\lambda|!}\right)^2\left|\frac{t}{L}\right|^{|\lambda|}\right]^2.
  \ebn
  Now according to the  Burnside's formula $\sum\nolimits_{\lambda\in\mathbb{Y},|\lambda|=n}\left(\dim\lambda\right)^2=n!$,
  the previous inequa\-lity can be rewritten as $\left|\mathcal{B}\left(\sigma,t\right)\right|\leq \exp\frac{2|t|}{L}$.

  The second assertion follows from the asymptotic behavior of the Barnes function coefficients in (\ref{irrcb1}).
  For example, one has
  \be\label{barnesas}
  \fl  G\left(1+2\left(\sigma+n\right)\right)G\left(1-2\left(\sigma+n\right)\right)=
   \frac{G\left(1-2\sigma\right)}{G\left(1+2\sigma\right)}\left(\frac{i\sin2\pi\sigma}{\pi}\right)^{2n}
   G^2\left(1+2\left(\sigma+n\right)\right).
  \eb
  The $n\rightarrow\infty$ asymptotics
  \ben
  \ln G\left(1+2\left(\sigma+n\right)\right)=2n^2\ln n+O\left(n^2\right)
  \ebn
  clearly dominates the factors $t^{\left(\sigma+n\right)^2}$ and $\displaystyle\frac{\sin^{2n}2\pi\sigma}{\pi^{2n}}$ in
  (\ref{irrcb1}) and (\ref{barnesas}). Together with the previous bound on $\left|\mathcal{B}\left(\sigma,t\right)\right|$,
  this suffices to complete the proof.
  \epf

  In the Painlev\'e VI case, a slight modification of the above argument shows that  the corresponding  series have non-zero radius of convergence. The related issues will be discussed elsewhere.

  \subsection{Long-distance expansion}
  The formal long-distance ($r\rightarrow \infty$) expansion of $u(r)$
  has the following form:
  \be\label{lde}
  u(r)=\sum_{k,l=0}^{\infty}\sum_{\epsilon=\pm}b^{\epsilon}_{kl}e^{i\epsilon(2k+1)r}r^{\left(2k+1\right)\left(i\epsilon \nu-\frac12\right)-l}+2\pi n,\qquad n\in\Zb,
  \eb
  where $ \Im \nu\in(-1,1)$ and $r\in\Rb_{>0}$. The expansion coefficients and the parameter $\nu$ can be recursively determined from the
  sine-Gordon equation in terms of the first two coefficients $b_{00}^{\pm}$ which can be arbitrary.
  In particular, one finds that
  \begin{eqnarray*}
  \nu=-\frac{b_{00}^+b_{00}^-}{4},\\
  b^{\pm}_{10}=-\frac{{b^{\pm}_{00}}^3}{2^4\cdot 3},\qquad
  b^{\pm}_{20}=\frac{{b^{\pm}_{00}}^5}{2^8\cdot 5},\qquad
  b^{\pm}_{30}=-\frac{{b^{\pm}_{00}}^7}{2^{12}\cdot 7},\\
  b^{\pm}_{01}=\pm\frac{i b^{\pm}_{00}}{8}\left(6\nu^2\pm 4i\nu-1\right),\\
  b^{\pm}_{11}=\pm\frac{9i b^{\pm}_{10}}{8}\left(2\nu^2\pm 2i\nu-1\right),\\
  b^{\pm}_{21}=\pm\frac{15i b^{\pm}_{20}}{8}\left(2\nu^2\pm 2i\nu-1\right),\\
  b^{\pm}_{02}=-\frac{b^{\pm}_{00}}{128}\left(36\nu^4\pm128i\nu^3-104\nu^2\mp 56i\nu+9\right),\\
  b^{\pm}_{12}=-\frac{3b^{\pm}_{10}}{128}\left(108\nu^4\pm296i\nu^3-336\nu^2\mp 236i\nu+71\right),\\
  b^{\pm}_{03}=\mp\frac{ib^{\pm}_{00}}{1024}\left(72\nu^6\pm 624i\nu^5-1788\nu^4\mp 1824i\nu^3+1522\nu^2
  \pm 532i\nu-75\right).
  \end{eqnarray*}
  The expression of the leading coefficients $b_{00}^{\pm}$ in terms of
  monodromy data $\left(\sigma,\eta\right)$  was found in \cite{IN,K1,Nov}, see also \cite[Chapter 14]{FIKN}. It reads
  \begin{eqnarray}\label{ldmndr1}
  e^{\pi\nu}=\frac{\sin2\pi\eta}{\sin2\pi\sigma},\\
  \label{ldmndr2}
  b^{\pm}_{00}=-e^{\frac{\pi\nu}{2}\mp\frac{i\pi}{4}}2^{1\pm 2i\nu}
  \frac{\Gamma\left(1\mp i\nu\right)}{\sqrt{2\pi}}\frac{\sin2\pi\left(\sigma\mp \eta\right)}{\sin2\pi\eta}.
  \end{eqnarray}

  The long-distance expansion of the tau function (somewhat surprisingly) exhibits
  a lot more structure than the series for $u(r)$, and looks
  similar to the short-distance series (\ref{tauexp0}).
  Using the relation (\ref{taudef}) and the above first coefficients of the expansion (\ref{lde}),
  one obtains
   \begin{eqnarray}\label{tauper}
 \fl &\tau \left(2^{-12}r^4\right)=\mathrm{const}\cdot r^{\frac14}\,e^{\frac{r^2}{16}}\,\Biggl\{\,
 r^{\frac{\nu^2}{2}}e^{\nu r}\left[1+\frac{\nu\left(2\nu^2+1\right)}{8r}+
 \frac{\nu^2\left(4\nu^4-16\nu^2-11\right)}{128r^2}\right]+\biggr.\\
 \fl \nonumber &+\frac{ib_{00}^+ }{4}r^{\frac{(\nu+i)^2}{2}}e^{(\nu+i) r}\left[1+
 \frac{\left(\nu+i\right)\left(2\left(\nu+i\right)^2+1\right)}{8r}+
 \frac{\left(\nu+i\right)^2\left(4\left(\nu+i\right)^4-16\left(\nu+i\right)^2-11\right)}{128r^2}
 \right]\\
 \fl \nonumber &+\frac{i b_{00}^- }{4}r^{\frac{(\nu-i)^2}{2}}e^{(\nu-i) r}\left[1+
 \frac{\left(\nu-i\right)\left(2\left(\nu-i\right)^2+1\right)}{8r}+
 \frac{\left(\nu-i\right)^2\left(4\left(\nu-i\right)^4-16\left(\nu-i\right)^2-11\right)}{128r^2}
 \right]\\
 \fl\nonumber &-\frac{{b_{00}^+}^2\left(\nu+i\right)}{64}\, r^{\frac{(\nu+2i)^2}{2}}e^{(\nu+2i) r}\left[1+
 \frac{\left(\nu+2i\right)\left(2\left(\nu+2i\right)^2+1\right)}{8r}\right]\\
 \fl\nonumber &-\frac{{b_{00}^-}^2\left(\nu-i\right)}{64}\, r^{\frac{(\nu-2i)^2}{2}}e^{(\nu-2i) r}\left[1+
 \frac{\left(\nu-2i\right)\left(2\left(\nu-2i\right)^2+1\right)}{8r}\right]
 +O\left(r^{\frac{\nu^2}{2}-3}e^{\nu r}\right)\Biggr\}.
 \end{eqnarray}
 The manifest periodic pattern leads us to the following conjecture, cf (\ref{tauexp0}):
 \begin{conj}\label{LDc}
 Long-distance expansion of the sine-Gordon/Painlev\'e~III$_3$ tau function is given by a convergent series
 \begin{eqnarray}
 \label{tauexpi1}
 \tau\left(2^{-12}r^4\right)=\chi\left(\sigma,\nu\right)\sum_{n\in\mathbb{Z}}e^{4\pi i n \rho}
 \mathcal{G}\left(\nu+in,r\right),\\
 \label{tauexpi2}
 \mathcal{G}\left(\nu,r\right)=e^{\frac{i\pi\nu^2}{4}}{2}^{\nu^2}\left(2\pi\right)^{-\frac{i\nu}{2}}G\left(1+i\nu\right)
 r^{\frac{\nu^2}{2}+\frac14}e^{\frac{r^2}{16}+\nu r}
 \mathcal{D}\left(\nu,r\right), 
 \end{eqnarray}
 where $\mathcal{D}\left(\nu,r\right)$ admits the asymptotic expansion
 \be\label{tauexpi3}
  \mathcal{D}\left(\nu,r\right)\sim 1+\sum_{k=1}^{\infty}D_k\left(\nu\right)r^{-k},\qquad r\rightarrow\infty.
 \eb
 In these formulae, $G(z)$ again stands for the Barnes $G$-function. The parameters $\left(\nu,\rho\right)$ are related to
 mo\-no\-dro\-my data by
 (\ref{ldmndr1}) and
 \be\label{rhonot}
  e^{4\pi i \rho}=\frac{\sin2\pi\eta}{\sin2\pi\left(\sigma+\eta\right)}.
  \eb
 \end{conj}
 Conjecture~\ref{LDc} is expected to hold for any  $\sigma,\eta\notin \mathbb{Z}/2$.
 The most straightforward way to test it is to recursively compute next terms in
 the asymptotic expansion of $u(r)$ and $\tau\left(2^{-12}r^4\right)$ using the sine-Gordon equation. On one hand, this determines the coefficients $D_{k}\left(\nu\right)$:
  \begin{eqnarray*}
 D_1\left(\nu\right)=\frac{\nu\left(2\nu^2+1\right)}{8},\\
 D_2\left(\nu\right)=\frac{\nu^2\left(4\nu^4-16\nu^2-11\right)}{128},\\
 D_3\left(\nu\right)=\frac{\nu\left(8\nu^8-108\nu^6+402\nu^4+269\nu^2-24\right)}{3\cdot 2^{10}},\\
 D_4\left(\nu\right)=\frac{\nu^2}{3\cdot 2^{12}}\left(2\nu^{10}-56\nu^8+585\nu^6-2326\nu^4
 -\frac{7831}{8}\nu^2+612\right),\\
 \ldots \ldots \ldots
 \end{eqnarray*}
 On the other hand, such a procedure should reproduce (and it does indeed!)
 the intriguing periodicity structure of (\ref{tauexpi1}).

 We have implicitly set up the tau function normalization
 by (\ref{tauexp0})--(\ref{irrcb2}).
 Therefore, the constant prefactor $\chi\left(\sigma,\nu\right)$ in (\ref{tauexpi1})
 can no longer be chosen arbitrarily --- in fact, it is completely determined by monodromy data. The crucial point is that the explicit form of  $\chi\left(\sigma,\nu\right)$ cannot be derived from the asymptotics  (\ref{lde}) and connection formulae (\ref{ldmndr1})--(\ref{ldmndr2})  alone. We address this problem in the following section.

  \section{Connection coefficient}
  The pairs $\left(\sigma,\eta\right)$ and
  $\left(\nu,\rho\right)$ were used to characterize the tau function behavior as $r\rightarrow 0$ and $r\rightarrow\infty$,
  respectively.
  Now it will become convenient to pick one parameter from each pair and use
  $\left(\sigma,\nu\right)$ as local coordinates on the space $\mathcal{M}$ of monodromy data.
  An important drawback of such labeling is that $\left(\sigma,\nu\right)$ does not fix $\eta$ (and hence monodromy) uniquely ($\mathrm{mod}\;\mathbb{Z}$): another solution of (\ref{ldmndr1}) would be given by $\frac12-\eta$. We will specify
  which of the two solutions is chosen whenever it may lead to confusion.

  Let us consider analytic continuation of the connection coefficient $\chi\left(\sigma,\nu\right)$ along a path
  in the space of monodromy data which joins the point $\left(\sigma,\nu;\eta\right)$ and
  $\left(\sigma+1,\nu;\eta\right)$ or $\left(\sigma,\nu+i;\eta\pm\frac12\right)$. The periodicity of the expansions
  (\ref{tauexp0}) and (\ref{tauexpi1}) implies that $\chi\left(\sigma,\nu\right)$ should satisfy two recurrence
  relations:
  \begin{eqnarray}
  \label{recr1}
  \frac{\chi\left(\sigma+1,\nu\right)}{\chi\left(\sigma,\nu\right)}=e^{-4\pi i \eta},\\
  \label{recr2}
  \,\frac{\chi\left(\sigma,\nu+i\right)}{\chi\left(\sigma,\nu\right)}\,\! =\,e^{4\pi i \rho}.
  \end{eqnarray}

  Before we proceed with the construction of the general solution of
  (\ref{recr1})--(\ref{recr2}), it is useful to make the following remark.
  \begin{prop}
  The pairs $\left(\sigma,\eta\right)$ and $\left(\nu,\rho\right)$, connected by (\ref{ldmndr1}) and (\ref{rhonot}),
  provide two sets of canonically conjugate local
  coordinates on the monodromy manifold $\mathcal{M}$. More precisely, the pull\-back of the symplectic form $\Omega=dv\wedge du$ under the inverse monodromy map  $\mathfrak{m}^{-1}$ 
  is given~by 
  \be\label{canon}
  \Omega^*=32\pi i\, d\eta\wedge d\sigma=32\pi  \, d\rho\wedge d\nu.
  \eb
  \end{prop}
  \pf 
  In the proof we shall follow the method which was used in \cite{BFT}
  for the evaluation of the KdV symplectic form in terms of the relevant scattering data --- the PDE analog of the monodromy data.
  
  The form $\Omega$ is preserved by the Hamiltonian flow. Hence we
  can calculate the expression for $\Omega^*$ in terms of $\left(\sigma,\eta\right)$ and $\left(\nu,\rho\right)$ 
  by considering the limits 
  \ben
  \Omega^*_0=\lim_{r\rightarrow 0} \Omega^*,\qquad
  \Omega^*_{\infty}=\lim_{r\rightarrow \infty} \Omega^*,
  \ebn
  respectively. For instance, it follows from the asymptotic expansion
  (\ref{shas2}) that
  \ben
  u(r\rightarrow 0)=\alpha \ln r+\beta + o(1),
  \qquad v(r\rightarrow 0)=\alpha+o(1),
  \ebn
  and therefore $\Omega^*=\Omega^*_0=d\alpha\wedge d\beta$.
  The first equality in (\ref{canon}) can now be obtained  by identifying $\alpha=2i(1-4\sigma)$ and $\frac{\partial \beta}{\partial\eta}\Bigl|_{\sigma}=4\pi$
  with the help of (\ref{shas}).
  Similarly using the long-distance expansion (\ref{lde}), one 
  obtains 
  \ben
  \Omega^*=\Omega^*_{\infty}=2i \,d b_{00}^+\wedge db_{00}^-.
  \ebn
   The second equality in (\ref{canon}) then follows from the relation
   $ b_{00}^+ b_{00}^-=-4\nu$ and the diffe\-rentia\-tion formula $\frac{\partial }{\partial\rho}\ln b_{00}^-\Bigl|_{\nu}=-4\pi i$.
  \epf
  
  It can be inferred from (\ref{canon}), or verified by straightforward differentiation of (\ref{ldmndr1}) and (\ref{rhonot}),  that
  \be\label{closed}
  \frac{\partial\eta}{\partial \nu}=i\frac{\partial\rho}{\partial\sigma}
  \,.
  \eb
  Furthermore, there exists a generating function $\mathcal{W}\left(\sigma,\nu\right)$
  of the canonical transformation between the two pairs such that
  \be\label{closed2}
  \eta=\frac{\partial\mathcal{W}}{\partial \sigma},\qquad \rho=-i\frac{\partial \mathcal{W}}{\partial\nu}.
  \eb
  \begin{prop}\label{GFp}
  The generating function 
  $\mathcal{W}\left(\sigma,\nu\right)$
   can be chosen as
  \begin{eqnarray}
  \fl\qquad 8\pi^2\mathcal{W}\left(\sigma,\nu\right)=
  \mathrm{Li}_2\left(-e^{2\pi i\left(\sigma+\eta-\frac{i\nu}{2}\right)}\right)+
  \mathrm{Li}_2\left(-e^{-2\pi i\left(\sigma+\eta+\frac{i\nu}{2}\right)}\right)
  -  \left(2\pi\eta\right)^2+\left(\pi\nu\right)^2,
  \label{GF}
  \end{eqnarray}
  where $\mathrm{Li}_2\left(z\right)$ denotes the classical dilogarithm and
  $\displaystyle\eta\left(\sigma,\nu\right)=\frac{\arcsin\left(e^{\pi\nu}\sin2\pi\sigma\right)}{2\pi}$.  The right side of (\ref{GF}) is understood as analytic continuation from an open neighbor\-hood of the domain $0<\eta<\sigma<\frac14$, $\nu\in\mathbb{R}_{<0}$, where the dilogarithms are defined by
  their principal branches.
  \end{prop}
  \pf 
  The differentiation formula $\mathrm{Li}_2'\left(z\right)=-z^{-1}\ln\left(1-z\right)$ implies that
  \begin{eqnarray*}
  \frac{\partial}{\partial\sigma}\biggl[\mathrm{Li}_2\left(-e^{2\pi i\left(\sigma+\eta-\frac{i\nu}{2}\right)}\right)+
  \mathrm{Li}_2\left(-e^{-2\pi i\left(\sigma+\eta+\frac{i\nu}{2}\right)}\right)\biggr]=\\ =-2\pi i\left(1+\frac{\partial\eta}{\partial \sigma}\right)\biggl[\ln\left(1+e^{2\pi i\left(\sigma+\eta-\frac{i\nu}{2}\right)}\right)-
  \ln\left(1+e^{-2\pi i\left(\sigma+\eta+\frac{i\nu}{2}\right)}\right)\biggr].
  \end{eqnarray*}
  On the other hand, from the easily verified identities
  \be\label{cosiden}
  2\cos\pi\left(\sigma+\eta\pm\frac{i\nu}{2}\right)=e^{i\pi\left(\pm\sigma\mp\eta-\frac{i\nu}{2}-4\rho\right)}
  \eb
  it follows that
  \ben
  \ln\left(1+e^{2\pi i\left(\sigma+\eta-\frac{i\nu}{2}\right)}\right)-
  \ln\left(1+e^{-2\pi i\left(\sigma+\eta+\frac{i\nu}{2}\right)}\right)=4\pi i \eta.
  \ebn
  Combining this with the previous relation, we immediately deduce that $ \eta=\frac{\partial\mathcal{W}}{\partial \sigma}$.
  The identity $\rho=-i\frac{\partial \mathcal{W}}{\partial\nu}$ is proven analogously.
  \epf

  Now consider the following 1-form on $\mathcal{M}$ (more precisely, on its ramified cover
  whose different sheets are associated with the pairs $\left(\sigma,\nu\right)$ projecting down to the same mono\-dromy):
  \ben
  \omega=\sigma d\eta+i\nu d\rho.
  \ebn
  The relation (\ref{closed}) implies that this 1-form is closed. It will be integrated along the
  paths of two types:
  \begin{enumerate}
  \item contours $\gamma_{\sigma}$ going from the point $\left(\sigma,\nu;\eta\right)$ to
  $\left(\sigma+1,\nu;\eta\;\mathrm{mod}\;\mathbb{Z}\right)$,
  \item contours $\gamma_{\nu}$ going from the point $\left(\sigma,\nu;\eta\right)$ to
  $\left(\sigma,\nu+i;\eta+\frac12\;\mathrm{mod}\;\mathbb{Z}\right)$
  \end{enumerate}
  The derivatives
  \begin{eqnarray*}
  \frac{\partial\eta}{\partial\sigma}=\cot2\pi\sigma\tan2\pi\eta,\\
  \frac{\partial\eta}{\partial\nu}=i\frac{\partial\rho}{\partial\sigma}=\frac{\tan2\pi\eta}{2},\\
  \frac{\partial\rho}{\partial\nu}=\frac{\sin2\pi\sigma}{4i\cos2\pi\eta\sin2\pi\left(\sigma+\eta\right)},
  \end{eqnarray*}
  are periodic under analytic continuation along the contours $\gamma_{\sigma,\nu}$. As a consequence, the
  function $\mathcal{A}\left(\sigma,\nu\right)$ defined by
  \be
  \mathcal{A}\left(\sigma,\nu\right)=-\int^{\left(\sigma,\nu\right)}\omega=\mathcal{W}\left(\sigma,\nu\right)-\sigma\eta-i\nu\rho
  \eb
  under appropriate prescription of the integration/analytic continuation contours (in the first and second expression, respectively) satisfies
  \begin{eqnarray*}
  \frac{\partial}{\partial\sigma}\biggl[\mathcal{A}\left(\sigma+1,\nu\right)-\mathcal{A}\left(\sigma,\nu\right)\biggr]
  =-\frac{\partial\eta}{\partial\sigma},\\
   \frac{\partial}{\partial\nu}\biggl[\mathcal{A}\left(\sigma+1,\nu\right)-\mathcal{A}\left(\sigma,\nu\right)\biggr]
  =-\frac{\partial\eta}{\partial\nu},\\
  \frac{\partial}{\partial\sigma}\biggl[\mathcal{A}\left(\sigma,\nu+i\right)-\mathcal{A}\left(\sigma,\nu\right)\biggr]
  \, =\,\frac{\partial\rho}{\partial\sigma},\\
  \frac{\partial}{\partial\nu}\biggl[\mathcal{A}\left(\sigma,\nu+i\right)-\mathcal{A}\left(\sigma,\nu\right)\biggr]
  \, =\,\frac{\partial\rho}{\partial\nu}.
  \end{eqnarray*}

  In other words, partial derivatives of $4\pi i\mathcal{A}\left(\sigma,\nu\right)$ satisfy the very same recurrence
  re\-lations as the partial derivatives of $\ln\chi\left(\sigma,\nu\right)$, and hence the difference of the two quantities can
  only be a function with periodic derivatives. The latter does not
  need to be periodic itself and can in principle be very complicated. However, already
  this first approximation provides an insight that suffices to
  solve (\ref{recr1})--(\ref{recr2}).

  Proposition~\ref{GFp} and the identity
  \ben
  \mathrm{Li}_2\left(e^{2\pi i z}\right)=-2\pi i \ln\hat{G}\left(z\right)-2\pi iz\ln\frac{\sin\pi z}{\pi}-
  \pi^2z\left(1-z\right)+\frac{\pi^2}{6},
  \ebn
  where $\displaystyle \hat{G}\left(z\right)=\frac{G\left(1+z\right)}{G\left(1-z\right)}$ and $z\in(0,1)$, imply that
  \ben
  4\pi i \mathcal{A}\left(\sigma,\nu\right)=\ln \frac{\hat{G}\left(\sigma+\eta+\frac{1-i\nu}{2}\right)}{
  \hat{G}\left(\sigma+\eta+\frac{1+i\nu}{2}\right)}+\begin{array}{c}\mathrm{elementary}\\ \mathrm{functions}\end{array}.
  \ebn
  Now we have the following result:
  \begin{prop}
  The general solution of (\ref{recr1})--(\ref{recr2}) is given by
  \begin{eqnarray}
  \nonumber\chi\left(\sigma,\nu;\eta\right)=\left(2\pi\right)^{i\nu}\exp\left\{ i\pi\left(\eta^2-2\sigma\eta
  -\sigma^2+\eta-\sigma-\frac{\nu^2}{4}\right)\right\}\times\\
  \label{solrec}
  \qquad\qquad \times
  \frac{\hat{G}\left(\sigma+\eta+\frac{1-i\nu}{2}\right)}{
  \hat{G}\left(\sigma+\eta+\frac{1+i\nu}{2}\right)}\, \chi_{\mathrm{per}}\left(\sigma,\nu;\eta\right),
  \end{eqnarray}
  where $\chi_{\mathrm{per}}\left(\sigma,\nu;\eta\right)$ is an arbitrary periodic function of $\sigma$ and $\nu$.
  \end{prop}
  \pf Let us denote $\mu=\eta+\sigma-\frac{i\nu}{2}$. Different possible choices of parameter values, as well as
   analytic continuation along $\gamma_{\sigma,\nu}$, can only shift $\mu$ by integers. Rewrite the prefactor in the right side of (\ref{solrec}) as
   \be\label{pref}
   \fl \left(2\pi\right)^{i\nu}
   \exp\left\{ i\pi\left(\mu\left(\mu+i\nu+1-4\sigma\right)+2\sigma^2-
   \frac{(\nu-i)(\nu+4i\sigma)}{2}\right)\right\}
   \frac{\hat{G}\left(\mu+\frac{1}{2}\right)}{
  \hat{G}\left(\mu+i\nu+\frac{1}{2}\right)}.
   \eb
   We claim that this expression is invariant under integer shifts of $\mu$. Indeed, it follows from
   the recurrence relation
   $\hat{G}\left(z+1\right)=-\pi\left(\sin\pi z\right)^{-1}\hat{G}\left(z\right)$ that the shift
   $\mu\mapsto\mu+1$ multiplies (\ref{pref}) by
   $\displaystyle e^{i\pi\left(2 \mu+i\nu-4\sigma\right)}\frac{\cos\pi\left(\mu+i\nu\right)}{\cos\pi\mu}$. The latter quantity
   is equal to $1$ thanks to the identities (\ref{cosiden}) used in the proof of Proposition~\ref{GFp}.

   Because of the last result the shift $\sigma\mapsto\sigma+1$ amounts to multiplication of (\ref{pref}) by
   $e^{i\pi\left(-4\mu+4\sigma-2i\nu\right)}=e^{-4\pi i \eta}$, which proves the functional relation (\ref{recr1}).
   To demonstrate the remaining relation (\ref{recr2}), it suffices to note that the shift $\nu\mapsto\nu+i$ produces an
   additional factor $\displaystyle \frac{e^{i\pi\left(-\mu+2\sigma-i\nu\right)}}{2\cos\pi\left(\mu+i\nu\right)}$,
   which is equal to $e^{4\pi i \rho}$ by (\ref{cosiden}).
  \epf

  Numerical experiments show that the unknown periodic function $\chi_{\mathrm{per}}\left(\sigma,\nu;\eta\right)$
  is in fact a constant. This constant can be determined with the help of an elementary solution of Painlev\'e~III$_3$ given by
  \be\label{sps}
  u(r)=0\;\mathrm{mod}\;2\pi\quad \Longleftrightarrow
  \quad\tau\left(2^{-12}r^4\right)=\mathrm{const}\cdot r^{\frac14}\,e^{r^2/{16}} .
  \eb
  The relevant monodromy parameters can be chosen as
  \ben
  \sigma=\eta=\frac14,\qquad \nu=0,\qquad \rho\rightarrow -i\infty.
  \ebn
  The connection coefficient computed directly from (\ref{sps}) is equal to
  \ben
  \chi\left(\text{\footnotesize $\frac14$},0\right)=e^{-\frac{i\pi}{8}}\chi_{\mathrm{per}}=\frac{2^{-\frac34}}{\sqrt{\pi}\;
  G^{2}\left(\frac12\right)}.
  \ebn
  Hence we finally arrive at
  \begin{conj}\label{cc}
  Connection coefficient $\chi\left(\sigma,\nu;\eta\right)$ for the Painlev\'e~III$_3$ tau function has
  the following expression in terms of monodromy data:
  \begin{eqnarray}
  \nonumber\chi\left(\sigma,\nu;\eta\right)=\left(2\pi\right)^{i\nu-\frac12}\exp\left\{ i\pi\left(\eta^2-2\sigma\eta
  -\sigma^2+\eta-\sigma-\frac{\nu^2}{4}+\frac18\right)\right\}\times\\
  \label{conn}
  \qquad\qquad\quad \times\,\frac{2^{-\frac14}}{
  G^{2}\left(\frac12\right)}\;
  \frac{\hat{G}\left(\sigma+\eta+\frac{1-i\nu}{2}\right)}{
  \hat{G}\left(\sigma+\eta+\frac{1+i\nu}{2}\right)},
  \end{eqnarray}
  where $\sigma$, $\eta$  and $\nu$ are related by (\ref{ldmndr1}).
  \end{conj}
  \noindent
  In addition to the functional relations (\ref{recr1})--(\ref{recr2}), our answer (\ref{conn})  satisfies
  \begin{itemize}
  \item
  periodicity property $\chi\left(\sigma,\nu;\eta+1\right)=\chi\left(\sigma,\nu;\eta\right)$,
  \item
  reflection symmetry $\chi\left(-\sigma,\nu;-\eta\right)=\chi\left(\sigma,\nu;\eta\right)$.
  \end{itemize}
  This reflects the corresponding symmetries of monodromy parameterization.
  One also has an interesting symmetry which relates connection coefficients associated
  to two different solutions of (\ref{ldmndr1}) with fixed $\sigma,\nu$:
  \be
  \chi\left(\sigma,\nu;\eta\right)\chi\left(\sigma,\nu;1/2-\eta\right)=
  \frac{\left(2\pi\right)^{i\nu-1}e^{-\frac{i\pi\nu^2}{2}}}{\sqrt{2}\,G^4\left(\frac12\right)\hat{G}\left(i\nu\right)}.
  \eb
  This is a Painlev\'e III$_3$ counterpart of an analogous result for Painlev\'e~VI connection coefficients,
  see \cite[formula (4.9)]{ILT}.

  We performed numerical tests of Conjecture~\ref{cc} 
  for random values of monodromy parameters.
  Fig.~1 provides an illustration of such checks. There we plot the real and imaginary part of
  the short-distance (blue curve) and long-distance (red curve) expansions of
  $e^{-\frac{r^2}{16}-\nu r}\tau\left(2^{-12}r^4\right)$. The monodromy parameters are chosen as
  \ben
  \left(\sigma,\nu;\eta\right)=\left(0.12-0.25i,0.34+0.29i;0.23+0.42i\right).
  \ebn
  We keep the terms up to $O\left(t^{\sigma^2+15}\right)$ in the short-distance expansion (\ref{tauexp0}) of $\tau(t)$
  and the terms up to $O\left(r^{\frac{\nu^2}{2}-\frac{15}{4}}e^{\frac{r^2}{16}+\nu r}\right)$ in the long-distance
  asymptotic  expansion (\ref{tauexpi1})--(\ref{tauexpi3}) of $\tau\left(2^{-12}r^4\right)$.

  \begin{figure}[!h]
 \begin{center}
 \resizebox{16cm}{!}{
 \includegraphics{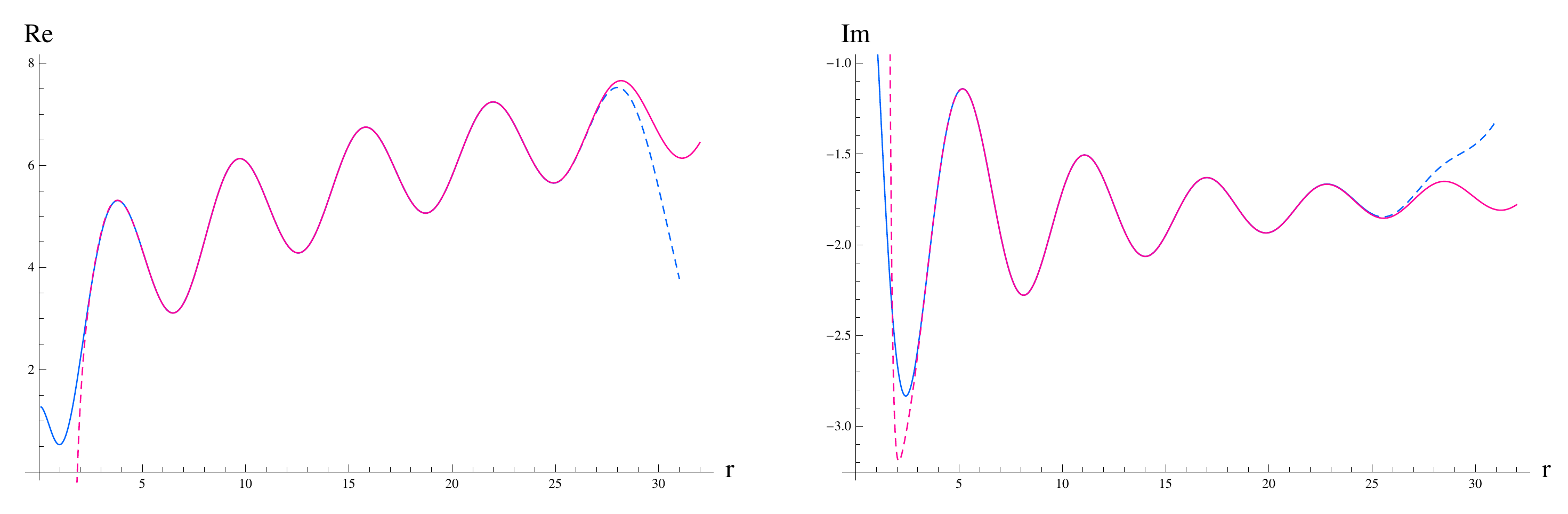}} \\
 Fig. 1. Short-distance (blue curve) vs long-distance (red curve) expansion.
 \end{center}
 \end{figure}

 \end{document}